\documentclass[twocolumn,showpacs,preprintnumbers,amsmath,amssymb]{revtex4}
\usepackage{amsmath}
\usepackage{amssymb}

\usepackage{graphicx}% Include figure files
\usepackage{dcolumn}% Align table columns on decimal point
\usepackage{bm}% bold math

\begin{document}

\title{Could a Classical Probability Theory Describe Quantum Systems?}
\author{Jinshan Wu$^{1}${\footnote{jinshanw@physics.ubc.ca}}, Shouyong Pei$^{2}$}
\affiliation{1. Department of Physics $\&$ Astronomy,\\
University of British Columbia, Vancouver, B.C. Canada, V6T 1Z1
\\ 2. Department of Physics, Beijing Normal University\\
Beijing, P.R. China, 100875}

\date{\today}

\begin{abstract}
Quantum Mechanics (QM) is a quantum probability theory based on the
density matrix. The possibility of applying classical probability
theory, which is based on the probability distribution
function(PDF), to describe quantum systems is investigated in this
work. In a sense this is also the question about the possibility of
a Hidden Variable Theory (HVT) of Quantum Mechanics. Unlike Bell's
inequality, which need to be checked experimentally, here HVT is
ruled out by theoretical consideration. The approach taken here is
to construct explicitly the most general HVT, which agrees with all
results from experiments on quantum systems (QS), and to check its
validity and acceptability. Our list of experimental facts of
quantum objects, which all quantum theories are required to respect,
includes facts on repeat quantum measurement. We show that it plays
an essential role at showing that it is very unlikely that a
classical theory can successfully reproduce all QS facts, even for a
single spin-$\frac{1}{2}$ object. We also examine and rule out
Bell's HVT and Bohm's HVT based on the same consideration.
\end{abstract}

\keywords{Quantum Mechanics, Hilbert Space, Probability Theory,
Quantum No-cloning Theorem, Hidden Variable Theory}

\pacs{03.67.-a, 03.65.Yz}

\maketitle

\section{The question and the common ground to start the discussion}

We regard quantum mechanics (QM), a theory based on wave amplitude
$\left|\phi\right>$ or density matrix $\rho$, as a quantum
probability theory (QPT) as it possesses the following properties:
first, for a given complete set of orthogonal vectors
$\left\{\left|\mu\right>\right\}$, it gives a classical probability
distribution $\left<\mu\right|\rho\left|\mu\right>$; second, for any
other such vector sets related with the former one by unitary
transformations, say $\left|\nu\right> =
\sum_{\mu}U_{\nu\mu}\left|\mu\right>$, it also gives another
classical probability theory, with probability distribution
$\left<\nu\right|\rho\left|\nu\right>$, which is related by the same
unitary transformations, $\left<\nu\right|\rho\left|\nu\right> =
\left<\mu\right|U^{\dag}\rho U\left|\mu\right>$. Meanwhile, what we
mean by classical probability theory (CPT), is a theory based on
the classical probability distribution function (PDF) instead of the density
matrix, possessing only the first property.

The goal of this work is to prove that QM could not be
described by a CPT. Or put in another way, QPT could never be
equivalently replaced by CPT. The question
is of the existence of a map from density matrix to probability
distribution function, and we want to show such map deos not exist.
Why we discuss this question, how we approach it, and
why we claim the answer is negative will be discussed. First we clarify our
terminology and establish an unambiguous language as the common
starting point of this discussion.

It is necessary to distinguish between the terms QM and quantum system (QS). By QM we
refer to the usual axiomatized system of quantum theory while
QS is reserved to refer to systems showing quantum properties in experiments.
Second, in this work, we limit our attention to quantum
measurement, excluding quantum evolution. Only axioms
about quantum measurement in QM and only quantum measurement
experiments are the subjects we will focus on. For example, if we
say CPT can describe QS, it means CPT can explain all
quantum measurement results. We wish to, for this moment, avoid the
discussion of CPT on evolution of QS because evolution
is less nontrivial but more technically intense. For example, we
would have to construct an equivalence of Schr\"{o}dinger's equation in
our CPT if we wanted to discuss evolution. Furthermore, in this work,
we deal only two systems, namely a single $\frac{1}{2}$-spin
system and an entangled two $\frac{1}{2}$-spin system. In QM
language, they both have finite dimension. Discussion on the above
systems can easily be generalized to general QS.

The QS are systems with the following
properties:
\newcounter{Lcount}
\begin{list}{QS-\Roman{Lcount}}
{\usecounter{Lcount}
    \setlength{\rightmargin}{\leftmargin}}
  \item There are a set of physical quantities associated with the system whose
  values we can measure. For each of them,
  when measurement is performed on a state of the quantum system,
  only finite outcomes will be observed. In addition, for every single measurement, only one
  specific outcome appears.
  \item If the same state is prepared, i.e. different realizations of the system go
  through the same preparation procedure, and the same measurement
  is performed on this ensemble, there is a statistical limit for the chance of appearance of a certain outcome.
  \item If in someway the quantum system is not
  destroyed and it can be measured again, then a repeat
  measurement of the same physical quantity will give us the same
  outcome as in the last measurement, with probability $1$.
  \item If a repeat measurement is made but of a different physical
  quantity, then still finite outcomes will be observed and their
  statistical limits also exist.
  \item The following property is given via a specific example of a spin-$\frac{1}{2}$
  system, but easily it can be generalized. Property of mixture state: the following
  two states can not be distinguished by any quantum measurements, including repeat
  measurements. Let's assume we have an apparatus preparing a spin into any desired states.
  Now state one is prepared as following: with probability $\frac{1}{4}$/$\frac{3}{4}$, we use
  the apparatus to prepare the spin into the $up$-/$down$- state along $z$ direction. State two is
  prepared with half possibility into the $up$-state along $\vec{r}_{1}=\left(\theta=\frac{2\pi}{3},
  \phi=\frac{\pi}{2}\right)$ direction
  and half possibility into the $up$-state along $\vec{r}_{2}=\left(\theta=\frac{2\pi}{3}, \phi=\frac{3\pi}{2}\right)$ direction.
\end{list}
Usual QM realize those properties of a QS through axioms:
\setcounter{Lcount}{1}
\begin{list}{QM-\Roman{Lcount}}
{\usecounter{Lcount}
    \setlength{\rightmargin}{\leftmargin}}
\item \label{QM1} States of a quantum object are normalized vectors in a complex linear
space $\mathcal{H}$ with dimension $N$, equipped with a definition
of inner product. Or equivalently, state of this object is described
by a $N\times N$ hermitian positive-defined normalized matrix
$\rho$. The set of such density matrices is denoted as
$\mathcal{N}\left(\mathcal{H}\right)$, the normalized positive operators
over $\mathcal{H}$.
\item \label{QM2}Physical quantities are hermitian operators over
$\mathcal{H}$. Their set is denoted as
$\mathcal{O}\left(\mathcal{H}\right)$. Physical quantities are
measurable. The measurement of $A$ on a system at state
$\rho$, results event $\alpha$ (meaning value of observable $A$ is
recorded as $\alpha$) with probability $p_{\alpha}$. $\alpha$
is one of the eigenvalues of $A$ (assumed non-degenerate but could be
trivially generalized) and $p_{\alpha} =
\left<\alpha\right|\rho\left|\alpha\right>$.
\item The state of the object after measurement, given the observed value is $\alpha$, is $\left|\alpha\right>\left<\alpha\right|$.
\end{list}

In a finite dimensional Hilbert space, the number of eigenvalues of an
operator is finite. QM-II realizes both QS-I and QS-II. QM-III
realizes QS-III in that if the same measurement is repeated, the outcome must be $\alpha$ and with probability $1$.

In order to realize QS-IV, one needs to consider basis transformations
in Hilbert space, that one vector could be expanded under difference
bases. After measurement of $A$ provided the outcome is event
$\alpha$, the system stays at
$\rho=\left|\alpha\right>\left<\alpha\right|$. If one then measures
for example $B$ with eigenvalues $\left\{\beta\right\}$, then according
to QM-II, the event of a specific $\beta$ will appear with probability $p_{\beta}=\left<\beta\right|\rho\left|\beta\right>
=\left<\beta\right|\left.\alpha\right>
\left<\alpha\right|\left.\beta\right>$. This explain QS-IV.

For, in QS-V, in usual QM language both preparations result the same
mixture state, $\rho
=\frac{1}{4}\left|\uparrow\right>\left<\uparrow\right| +
\frac{3}{4}\left|\downarrow\right>\left<\downarrow\right|$.
Therefore no measurement can tell their difference. And conceptually
we want our alternative theory, whatever it is, to respect the idea
that the mixture state, as in experiment of QS-V, is a probability
summation of all the exclusive possibilities. Our usual QM does
so. For example, we can confirm state two leads to the same density
matrix via the probability summation rule,
\begin{equation}\label{rho_QSV}
\left[\begin{array}{cc}\frac{1}{4} & 0 \\ 0 &
\frac{3}{4}\end{array}\right] =
\frac{1}{2}\left[\begin{array}{cc}\frac{1}{4} & i\frac{\sqrt{3}}{4}\\
-i\frac{\sqrt{3}}{4} & \frac{3}{4}\end{array}\right] +
\frac{1}{2}\left[\begin{array}{cc}\frac{1}{4} &
-i\frac{\sqrt{3}}{4}\\ i\frac{\sqrt{3}}{4} &
\frac{3}{4}\end{array}\right].
\end{equation}
Therefore QM realizes all the five experimental facts of a QS.

Next we construct a classical theory for all five
experimental facts. Such theory deos not need to respect the QM axioms at all, but it should still respect QS-I, QS-II, QS-III, QS-IV
and QS-V. In a usual discussion of HVT, only the first two are
required to be respected by the theory. We will see that if only
these two are required it is not impossible to have a classical
theory.

We have to mention that as an experimental fact, QS-III/IV is not
unquestionable. Usually the state is destroyed after measurement.
However, quantum nondemolition measurements\cite{nondemolition}
(QNM) allows a system to be subject to repeat measurements.
Therefore we take QS-III/IV also as an experimental fact. Another
thing worth mentioning is we did not include finite accuracy of real
measurements into our experimental facts. Our usual QM (QM-I, II,
III) embraces non-zero commutators between operators so it support
the idea of Uncertainty Principle. However, as argued by Bohm
\cite{bohm}, on the fundamental level one could not tell if it is
really impossible to measure some quantities simultaneously or it is
just because of problems on technology or accuracy of experiments.
This gives us the possibility to relax non-commutation relations
between physical quantities when necessary.

The question of the possibility of
fulfilling QS-I, II, III, IV and V by a CPT, requires to solidify explicitly what we refer to as a CPT. It
means the following:
\setcounter{Lcount}{1}
\begin{list}{CPT-\Roman{Lcount}}
{\usecounter{Lcount}
    \setlength{\rightmargin}{\leftmargin}}
  \item States form a set of event $\Omega$. There is a map
  $P$ from $\sigma$-Algebra $\mathcal{F}$ of $\Omega$ to
  $\left[0,1\right]$. $P$ satisfies the Kolmogorov axioms of probability\cite{Kolmogorov}.
  Only physical quantities corresponding to
  members of $\mathcal{F}$ are observable. A simpler case, which
  is quite often the case of a physical system, is that the set $\Omega$ is
  a set of countable simple events and $\mathcal{F}$ is the trivial topology, set of all
  subsets of $\Omega$. In our discussion, we only
  work with this simpler case. For exclusive events, if $A\cap B=\phi,
  A,B\in\mathcal{F}$, then $P\left(A\cup B\right)=P\left(A\right) +
  P\left(B\right)$. And for independent events, $A\otimes B\in\mathcal{F}\left(\Omega_{1}\otimes\Omega_{2}\right)$
  where $A\in\mathcal{F}\left(\Omega_{1}\right), B\in\mathcal{F}\left(\Omega_{2}\right)$, then $P\left(A\otimes
  B\right)=P\left(A\right)\cdot
  P\left(B\right)$.
  \item When the measurement of any $A\in \mathcal{F}$ is performed, every value of $\omega \in A \subseteq \Omega$ could be observed, with
  corresponding probability $P\left(\omega\right)$.
  \item After the measurement, the state of the system is the one
  observed. Provided event $\omega$ is recorded, the state of the object after
  measurement is $\omega$.
\end{list}
The validity of CPT-III is not really explicitly defined in the usual
probability theory. CPT itself provides no answer at all for that,
but usually people like to interpret CPT in this way. For
example, imagine a truly random perfect dice. After it is measured,
one would like to say it is at the state just observed by us.
However, experimentally what it really means is if the dice is
measured again, it guarantees one will observe the same value with
probability $1$. Therefore, although CPT-III is quite natural, it
can be altered if necessary.

After clarifying the terminology, now the question we seek to
discuss is better defined. We are looking for a CPT, which follows
CPT-I, CPT-II, CPT-III, of two systems: a single $\frac{1}{2}$-spin
and two entangled $\frac{1}{2}$-spins, which both possess QS-I,
QS-II, QS-III, QS-IV and QS-V. We will set the state of the
single-spin system at the $up$- state along the $x$ direction and
and the two-spin system at a singlet state. Although we aim at using
CPT as an alternative theory for quantum systems, we will still use
the usual QM language to denote their states. In another words, we
admit QM is a theory for quantum systems but we seek to determine if
QS can also be described by a classical theory such as CPT.

In section
$\S$\ref{language} we will put both CPT and QPT into a density
matrix form so that we can use the same mathematical language to
discuss the two theories. In section $\S$\ref{why} we discuss why we want to have such a map. Afterwards we
will present a CPT for a single-spin system and a CPT for a two-spin
system, in section $\S$\ref{sechidden} and
$\S$\ref{sechidden2} respectively. We will see that what
kind of CPT is necessary to fully describe quantum systems. We will then discuss why our CPT violates Bell's inequality and what is the
possible interpretation of such CPT. Finally in section
$\S$\ref{conclusion} we conclude that if we are willing to accept
all the prices we have to pay to have such a CPT for quantum
system, our CPT could be the one. But it is even harder to be
understood as compared with the usual QM.

\section{Density matrix language for both classical and quantum systems}
\label{language} In density matrix language for QM, the state of a
quantum object is represented by a density matrix
$\rho^{q}\left(t\right)$. The evolution is described by a unitary
transformation $U\left(t\right) \triangleq U\left(0,t\right)$ as
\begin{equation}
\rho^{q}\left(t\right) =
U\left(t\right)\rho^{q}\left(0\right)U^{\dag}\left(t\right),
\end{equation}
where generally $U\left(t\right)$ is determined by $H$, the
Hamiltonian of the quantum object. For a pure
initial state, the above density matrix formalism is equivalent
with the usual wave function or right vector formalism, but it can
also describe a mixture state. For example, we can consider an
exclusive mixture state as used in Von Neumann's picture of
quantum measurement\cite{von},
\begin{equation}
\rho^{q} =
\sum_{i}p_{i}\left|\phi_{i}\left>\right<\phi_{i}\right|,
\label{vonstate}
\end{equation}
where $\left\{\left|\phi_{i}\right>\right\}$ is a set of
orthogonal normalized vectors. According to Von Neumann's picture,
the meaning of such an exclusive state is that every sample of this
object chooses one of $\left\{\left|\phi_{i}\right>\right\}$ with probability $p_{i}$.

This explanation reminds us of the PDF of a truly random classical
object (TRCO), which generally should be included as objects of
classical mechanics (CM). A state of a TRCO is a PDF
$p\left(x\right)$ normalized over $\Omega = \left\{x\right\}$, the
set of all its possible states. It's therefore possible to
rewrite this PDF as a density matrix
\begin{equation}
\rho^{c} =
\sum_{x\in\Omega}p\left(x\right)\left|x\left>\right<x\right|,
\label{wustate}
\end{equation}
which gives exactly the same
information provided by a PDF. One could just
regard this as another notation of a discrete PDF. For the purpose
of normalization we also require that simple events are exclusive,
\begin{equation}
\left<x\left|\right.x^{\prime}\right> =
\delta\left(x-x^{\prime}\right),
\end{equation}
where $\delta\left(x-x^{\prime}\right)$ is the Kronecker delta for
the discrete set $\Omega$. In QM, generally a state of a
quantum system is a full structure density matrix, while in CM, a
state of a TRCO is a diagonal density matrix. Considering only
discrete sets allow us to use the notation $\left|x\right>$ and inner
product $\left<x\right.\left|y\right> = \delta_{xy}$ without any
problem. Although physicists also use such notation for continuous
systems, mathematicians do not like the idea of using
$\left|x\right>$ as a basis vector, or using Dirac $\delta$ function as
a basis of function space, for continuous system. Most expressions in
physicists' notation can be mapped onto more rigorous
mathematicians' notation\cite{Ballentine}, but we don't wish to deal with that
here. We limit our description to discrete
systems only. For example, a perfect dice is such an object.

From this point foward, we are going to use density matrix notation for both
QS and TRCO. Furthermore, we can construct a
similar theory to describe classical evolution processes. For
example, if we denote the evolution process as a linear operator
$\mathcal{T}$, then time evolution of such classical objects can
be defined as
\begin{equation}
\rho^{c}\left(t\right) \triangleq
\mathcal{T}\left(\rho^{c}\left(0\right)\right) = \sum_{x}
p\left(x\right)\mathcal{T}\left(\left|x\right>\left<x\right|\right).
\label{cm}
\end{equation}
Formally, we can use the evolution operator $T$ as $
\mathcal{T}\left(\left|x\right>\left<x\right|\right) = \left(T
\left|x\right>\right)\left(\left<x\right|T^{\dag}\right) =
\left|x\left(t\right)\right>\left<x\left(t\right)\right|$, so
\begin{equation}
\rho^{c}\left(t\right) =T \rho^{c}\left(0\right)T^{\dag}.
\end{equation}
Also $TT^{\dag} = T^{\dag}T = I$, which can be proved as follows:
first, for system fully determined by $x$, we have
$\delta\left(x\left(t\right)-y\left(t\right)\right)=
\delta\left(x-y\right)=\left<x\left|\right.y\right>$, then
\begin{equation}
\left<x\right|T^{\dag}T \left|y\right> =
\left<x\left(t\right)\left|\right.y\left(t\right)\right> =
\delta\left(x\left(t\right)-y\left(t\right)\right).
\end{equation}
Therefore, both QM and CPT are unitary evolution theories of density
matrices, while the difference between them is the existence of
off-diagonal elements. From this point of view, our task in this
work is to put a full-structure density matrix into a diagonal
density matrix. We call this a question of
finding a diagonalization map. The reason we introduce TRCOs is to help towards the
understanding of classical objects and, later, quantum objects. By
emphasizing ``truly random'', we are not refering to objects which behave
randomly because of the uncertainty in their initial conditions. For a
TRCO there is intrinsically no way, even for ``God'', to tell its
real state before a measurement is performed. We only can say it
stays in a classical mixture state. One may argue that a physical
classical object is not a truly random object. Imagining such a TRCO, however, will help us to understand the
classical and quantum measurement process.

To conclude this section, we want to point out that our language of
the diagonal density matrix for CPT and the non-diagonal density matrix for
QM provides a unified description of classical and quantum
mechanics. Besides this, there is another set of language based on
$C^{*}$-algebra\cite{C*algebra} also unifying the description of
classical and quantum mechanics. There, classical and quantum
operators are more basic descriptions of a system and they form
an abelian $C^{*}$-algebra and a non-abelian
$C^{*}$-algebra respectively. States are defined as functionals over the
corresponding algebra. Although we will not prove it explicitly
here, we believe that our notation of the diagonal and non-diagonal
density matrix for classical and quantum systems, is in fact
equivalent with the $C^{*}$-algebra based language.

\section{Why are we looking for such a map?}
\label{why}

If we have a CPT as desired, it is a HVT of quantum systems.
Quantum systems are no longer quantum but TRCOs. Therefore, one can understand quantum measurement if one can understand measurement of TRCOs.
The most straightforward picture of a measurement is a measurement
on a determinant classical object. Assumed as a discrete system, it
stays in state $\left|x\right>\left<x\right|$ before it is
measured. After the measurement we get the information that it was
in state $\left|x\right>\left<x\right|$ and it remains in state $\left|x\right>\left<x\right|$. The less straightforward picture of a measurement is a measurement on a
statistically random classical object. Here the term ``statistically
random'' means that the nature of this object is still determinant,
but with incomplete information it appears as a random object. For
every given such object, we just do not know its state but it is
already fixed. This also means its randomness is only meaningful as
in an ensemble. This is called statistical interpretation
of probability theory. Again for such an object, it is in state
$\left|x\right>\left<x\right|$ before measurement and after the
measurement we get the information that it was in state
$\left|x\right>\left<x\right|$ and it remains in state
$\left|x\right>\left<x\right|$. Notice that although $x$ can be
one of a large set, but it is fixed with probability
$p\left(x\right)$ before the measurement is performed.

Measurement on a TRCO is less understandable. Assuming
such an object really exists for the moment, its state is unknown
before measurement. After the measurement we find with probability
$p\left(x\right)$ that it was in state
$\left|x\right>\left<x\right|$ and it remains
in state $\left|x\right>\left<x\right|$ afterwards. Here we find that a
phenomena so-called ''collapse'' of probability function has occurred.
While this seems less understandable, both ``statistical
randomness'' and ``true randomness'' give us the same measurement
result. One could never distinguish which is the ``real'' one from
measurements. It is a philosophical question to ask which one
it really is, statistically random or truly random and so from now on
we will treat them as the same.

Now imagine we have two correlated TRCOs which have
exactly the same states, but unknown. Since they are both TRCOs we
do not know their states before the measurement. If we measure one of
them, say we find that it stays in state
$\left|x\right>\left<x\right|$, then we immediately know the state
of the other object is also $\left|x\right>\left<x\right|$. In this sense, if we assume the existence of such TRCOs, ``spooky
action'' exists even in classical mechanics. A classical bit of information need to be transfered from one to the other in order for the other to know that its counterpart's state after the measurement. Quantum ``spooky
action'' in entangled systems is not stranger than its classical
version at all. They are different just that in the quatnum case, both direction and measurement outcome need to be transfered, not only the outcome.

Provided there is a TRCO description of quantum systems,
the two problems of quantum measurement, namely collapse of the wave
function and measurement of entangled states, become collapse of
the probability function and measurement of classical correlated states
in measurement of TRCOs. This implies that if one believes
measurement of TRCOs is understandable, then measurement of quantum
systems is also understandable.

Here we assume TRCO Assumption: there is no difficulty or
confusion in understanding measurement of TRCOs. Even if it is questionable, if TRCO can describe quantum systems,
then we know the problem of quantum measurement comes from classical probability theory and has nothing to do with
any other quantum nature. Of course the situation will be different if
we find out that TRCOs can not describe quantum systems.

We can formally compare measurement of TRCOs and quantum systems.
Here we include both auxiliary system $m$ and object system $o$
explicitly into our formal description. The measurement includes two
steps. First, a classical correlated state is formed by an interaction
process, so that from an initial state
\begin{equation}
\rho^{c,o} = \sum_{x} p\left(x\right)\left|x\right>\left<x\right|.
\label{classicalstate}
\end{equation}
we get
\begin{equation}
\rho^{c,o}\otimes\rho^{c,m}\longrightarrow\rho^{c,om} = \sum_{x}
p\left(x\right)\left|x\otimes M\left(x\right)\right>\left<x\otimes
M\left(x\right)\right|. \label{classicalmeasure}
\end{equation}
Second, when we only check the value recorded on the auxiliary
system, we get a sample from the auxiliary system's partial
distribution, which is
\begin{equation}
\rho^{c,m}\triangleq tr^{o}\left(\rho^{c,om}\right) = \sum_{x}
p\left(x\right)\left|M\left(x\right)\right>\left<
M\left(x\right)\right|, \label{classicalmeasurefinal}
\end{equation}
where $tr^{o}$ means the trace is taken over object state space,
a standard procedure in probability theory when only information
on the partial distribution is needed. Therefore, according to their
exclusiveness nature and CPT-III, the sampling process gives us one
specific state $M\left(x^{*}\right)$. This happens with the
desired probability $p\left(x^{*}\right)$, due to CPT-II.
$M\left(x^{*}\right)$ on the auxiliary system means
$x^{*}$ on the measured object.

However, even formulated in the same way but in usual QM language,
the picture of quantum measurement is different because
the general quantum density matrix has non-zero off-diagonal
terms. As in equ(\ref{classicalmeasure}) and
equ(\ref{classicalmeasurefinal}), with first an interacting process and
then a partial trace, if the same steps are applied onto a quantum
system with
\begin{equation}
\rho^{q,o}
=\sum_{\mu\nu}\rho_{\mu\nu}\left|\mu\right>\left<\nu\right|,
\label{quantumstate}
\end{equation}
then, firstly,
\begin{equation}
\rho^{q,o}\otimes\rho^{q,m}\longrightarrow\rho^{q,om} =
\sum_{\mu\nu} \rho_{\mu\nu}\left|\mu\otimes
M\left(\mu\right)\right>\left<\nu\otimes M\left(\nu\right)\right|,
\label{quantummeasure}
\end{equation}
and secondly, when we only check the value recorded on the auxiliary
system, we obtain a sample from the auxiliary system's partial
distribution, which is a sample of
\begin{equation}
\rho^{q,m}\triangleq tr^{o}\left(\rho^{q,om}\right) =
\sum_{\lambda}
\rho_{\lambda\lambda}\left|M\left(\lambda\right)\right>\left<
M\left(\lambda\right)\right|. \label{quantummeasurefinal}
\end{equation}
However, for a quantum object, equ(\ref{quantummeasurefinal}) is not
a copy of equ(\ref{quantumstate}), while
equ(\ref{classicalmeasurefinal}) is an exact copy of
equ(\ref{classicalstate}) for a classical object. Therefore, if TRCOs
could never describe quantum system, even with the TRCO Assumption,
quantum measurement is still harder to understand than measurement
of TRCOs. However, if we have a CPT for quantum system, then quantum
measurement is just as understandable as measurement of a TRCO.

As we have seen, due to CPT-III, a classical measurement ends up
with an exact copy of the object state. We may regard such a process
as a clone. However, this clone does not respect the definition of
clone in the original quantum non-cloning theorem\cite{clone} (QNCT),
\begin{equation}
\rho^{o}_{aim}\otimes\rho^{m}_{initial}
\overset{U}{\longrightarrow}
\rho^{om}=\rho^{o}_{aim}\otimes\rho^{m}_{aim}, \label{nonclone}
\end{equation}
while now it has more general property
$\rho^{o}_{aim}\otimes\rho^{m}_{initial}
\overset{U}{\longrightarrow} \rho^{om}$ that,
\begin{equation}
tr^{m}\left(\rho^{om}\right) = \rho^{o}_{aim} \text{ and }
tr^{o}\left(\rho^{om}\right) = \rho^{m}_{aim}, \label{clone}
\end{equation}
Equ(\ref{nonclone}) is a special case of equ(\ref{clone}). In fact,
this more general clone is called a broadcast and it has been proved
that a quantum system can not be broadcasted in quantum
no-broadcasting theorem (QNBT)\cite{broadcast}. Unless the object
system initially stays in one of a set of {\bf{known}} orthogonal
states, a quantum system can not be broadcasted. In our language,
this means when a system is in a classical probability combination
of known orthogonal states, i.e a diagonal density matrix under a known
basis, it can be broadcasted. This is just a broadcast of TRCOs.

An arbitrary unknown state of a TRCO can be broadcasted, or a
diagonal density matrix state can be broadcasted. Therefore, if the
above diagonalization mapping exists, through it, a quantum
system can also be broadcasted. This would conflict with QNBT, which is proved in the language of usual QM.
This leads to two possibilities: firstly, QNBT holds and
diagonalization mapping does not exist; or secondly, QNBT is not
valid and the mapping exists. Now we find that QNBT is also reduced
to the existence of the diagonalization mapping. Therefore, it seems
all the confusing and ``extraordinary'' problems in QM including
quantum measurement, HVT and QNBT come down to one question, the
existence of such diagonalization mapping.

The relation between QNBT and HVT can be shown more explicitly. A
TRCO can be broadcasted, by introducing a classical hidden variable.
For example, let us use a perfect two-face dice as a TRCO. We
introduce a classical signal $\lambda$, generated from a given
PDF $\rho\left(\lambda\right)$ over $\Gamma=
\left\{\lambda\right\}$. The state of the dice is determined by this
signal as follows,
\begin{equation}
\rho^{c,o} = \sum_{\lambda \in
\Gamma}\rho_{+}\left(\lambda\right)\left|+\right>\left<+\right| +
 \sum_{\lambda \in
\Gamma}\rho_{-}\left(\lambda\right)\left|-\right>\left<-\right|,
\end{equation}
where it is required that
\begin{equation}
\int_{\Gamma}d\lambda \rho_{+}\left(\lambda\right) = \frac{1}{2}
=\int_{\Gamma}d\lambda \rho_{-}\left(\lambda\right).
\end{equation}
We then duplicate this hidden variable signal, send a copy to
another dice while the original signal is sent to the original dice.
Each dice determines its state respectively according to
the value of its hidden variable. Now we get a broadcast of the
dice. In this sense, it is fair enough to say that the success of an
HVT for CM makes it possible to broadcast a classical object. So what
about a HVT for QM?

\section{A possible TRCO and understanding of its measurement}
Consider a quantum system coupled with a large thermal bath whose
eigenenergy can be measured in much shorter time than the relaxation time. Our
measurement is performed once in a while with the time interval between measuremens being much
longer than the typical relaxation time of this system. The outcomes of such measurements
will give us a sequence of eigenvalues whose probability of appearance follows classical Boltzmann distribution.
Do we now believe that the system stays in one of the eigenstates
before any measurements? And further, does our belief matter? It
seems there is no difficulty in accepting the results from
this measurement as is. From this example, we wish to argue
that our assumption of the existence of TRCO and validity of TRCO
Assumption, which states there is no problem in understanding
measurement of TRCOs, is plausible.

\section{CPT for single-spin system}
\label{sechidden}

The possibility of a CPT or a HVT for quantum system has been long
investigated by many great physicists\cite{bell,
bellrmp, bohm, bohmrmp, scally}. Bell's Theorem\cite{bell} says that all local HVT should obey the Bell's inequality,
which is not respected by QM. Experimental tests suggests that QS
do violate the Bell's inequality so QM is a preferred theory for
QS\cite{belltest}. But this statement has not yet been supported by
all physicists. In the following, we will try to answer this problem
in another way. We are willing to go as far as possible to construct
a CPT to give consistent results with quantum systems including QS-I,
II, III, IV and V. If this effort fails we will find where and why;
or if it succeeds, we will check whether it is acceptable or not.
If it succeeds, according to Bell's Theorem, it should
be non-local. It will be interesting to show explicitly the place
where non-locality enters the theory. In fact, in \cite{scally}, the
author already discussed a similar question of ``How to make quantum mechanics look like a hidden-variable theory and vice versa'' using the Wigner distribution. Here
in this paper, to discuss the same question, we start from a more
general form of CPT and try to make it successful as far as possible. For
simplicity of language, in this paper, we regard CPT and HVT of
a quantum system as being the same meaning and later on just simply call
them HVT.

According to our general framework, HVT could be in a
classical diagonal density matrix form,
\begin{equation}
\rho^{hvt} =
\sum_{\lambda\in\Gamma}\rho\left(x\left(\lambda\right)\right)
\left|x\left(\lambda\right)\right>\left<x\left(\lambda\right)\right|,
\label{hvtstate}
\end{equation}
where $x$ is the dynamical variable, $\lambda$ is the hidden
random variable and $x\left(\lambda\right)$ is an onto mapping,
$\rho\left(x\left(\lambda\right)\right)$ is a PDF over $\Gamma =
\left\{\lambda\right\}$, a set of exclusive events,
\begin{equation}
\left<x\left(\lambda\right)\right|\left.x\left(\lambda^{\prime}\right)\right>
= \delta\left(\lambda-\lambda^{\prime}\right).
\label{hvtcondition}
\end{equation}
One thing that is necessary to be pointed out is here the parameter
$\lambda$ is abstract, not limited as a single variable. A
successful HVT has to respect all QS facts. We will start from QS-I
and II.

\subsection{CPT based on exclusiveness of all elementary pure events}
We first consider a single spin-$\frac{1}{2}$ as in Bohm's
HVT\cite{bohm}, and then focus on an entangled object with two
subsystems as discussed in Bell's inequality\cite{bell}. For
simplicity, let's just consider a specific quantum state, a
$\frac{1}{2}$-spin in the state of $\left|\uparrow\right>_{x}$, the
$up$ state of $S_{x}$. In the language of QM, it's
\begin{equation}
\rho^{q} =
\frac{1}{2}\left(\left|\uparrow\right>_{z}\left<\uparrow\right|_{z}
+ \left|\uparrow\right>_{z}\left<\downarrow\right|_{z} +
\left|\downarrow\right>_{z}\left<\uparrow\right|_{z} +
\left|\downarrow\right>_{z}\left<\downarrow\right|_{z}\right).
\end{equation}
For a HVT, the first trial density matrix will naturally be,
\begin{equation}
\rho^{hvt} = \sum_{\lambda_{z}}
\rho_{+}\left(\lambda_{z}\right)\left|\uparrow\right>_{z}\left<\uparrow\right|_{z}
+ \sum_{\lambda_{z}}
\rho_{-}\left(\lambda_{z}\right)\left|\downarrow\right>_{z}\left<\downarrow\right|_{z},
\label{hvtrho}
\end{equation}
with the following requirement to give correct results for
measurement on $S_{z}$,
\begin{equation}
\int_{\Gamma_{z}} d\lambda_{z}\rho_{+}\left(\lambda_{z}\right) =
\frac{1}{2} = \int_{\Gamma_{z}}
d\lambda_{z}\rho_{-}\left(\lambda_{z}\right). \label{require}
\end{equation}
However, this gives the consistent results with QS-I and II only for
$S_{z}$ measurement. We can also measure $S_{x}$. If we still
respect the possible non-commutative relation between quantum
operators $S_{x}$ and $S_{z}$, then we need to do a basis
transformation in $\mathcal{H}^{q}$ and do measurement of $S_{x}$.
We get
\begin{equation}
\begin{array}{ccc}
\rho^{hvt} & = & \frac{1}{2}\sum_{\Gamma_{z}}
\left[\rho_{+}\left(\lambda_{z}\right) +
\rho_{-}\left(\lambda_{z}\right)\right]\left(\left|\uparrow\right>_{x}\left<\uparrow\right|_{x}
+ \left|\downarrow\right>_{x}\left<\downarrow\right|_{x}\right) \\
&&+ \frac{1}{2}\sum_{\Gamma_{z}}
\left[\rho_{+}\left(\lambda_{z}\right) -
\rho_{-}\left(\lambda_{z}\right)\right]\left(\left|\uparrow\right>_{x}\left<\downarrow\right|_{x}
+ \left|\downarrow\right>_{x}\left<\uparrow\right|_{x}\right).
\end{array}
\end{equation}
We can see that, according to equ(\ref{require}), the result
of this measurement will be $\frac{1}{2}$ probability to get $up$
and $\frac{1}{2}$ to get $down$. This is obviously wrong. We
know for the specific state we choose above, the correct result of
the $S_{x}$ measurement is the $up$ state only. This HVT does not
realize QS-I and II.

There is one way to overcome this inconsistency with the price that not one hidden variable, but another hidden
variable is needed. In order to get correct results for measurement
on $S_{z}$ and $S_{x}$, we need
\begin{align}
\notag \rho^{hvt} = \frac{1}{\mathcal{N}}\left[ \sum_{\lambda_{z}}
\rho_{+}\left(\lambda_{z}\right)\left|\uparrow\right>_{z}\left<\uparrow\right|_{z}
+ \sum_{\lambda_{z}}
\rho_{-}\left(\lambda_{z}\right)\left|\downarrow\right>_{z}\left<\downarrow\right|_{z}\right.
\\
\left.+\sum_{\lambda_{x}}\rho_{+}\left(\lambda_{x}\right)\left|\uparrow\right>_{x}\left<\uparrow\right|_{x}
+
\sum_{\lambda_{x}}\rho_{-}\left(\lambda_{x}\right)\left|\downarrow\right>_{x}\left<\downarrow\right|_{x}\right],
\end{align}
with the requirement,
\begin{equation}
\int_{\Gamma_{x}} d\lambda_{x}\rho_{+}\left(\lambda_{x}\right) = 1,
\int_{\Gamma_{x}} d\lambda_{x}\rho_{-}\left(\lambda_{x}\right)=0.
\label{disaster}
\end{equation}
$\mathcal{N}$ is a normalization constant to keep
$tr\left(\rho\right)=1$ and here $\mathcal{N}=2$. With this density
matrix, a measurement of $S_{x}$ will give the $up$ state only. We can
similarly include $S_{y}$ terms using another hidden variable
$\lambda_{y}$. However, a successful HVT should respect QS-I and II
for measurement on an arbitrary direction. For this purpose, will
three hidden variables corresponding to $S_{x},S_{y},S_{z}$ be
enough? For example, for a measurement of
\begin{equation}
S_{r} = \sin{\theta}\cos{\phi}S_{x}+ \sin{\theta}\sin{\phi}S_{y} +
\cos{\theta}S_{z}, \label{inherent}
\end{equation}
on the above state, the possible outcomes are
\begin{equation}
s_{r} = \frac{1}{2}\left(\sin{\theta}\cos{\phi} \pm
\sin{\theta}\sin{\phi} \pm \cos{\theta}\right).
\end{equation}
This could be continuous number, not only $\pm\frac{1}{2}$. We see
that it does not respect QS-I. So QS-I requires one hidden variable
for measurement on every direction and abandonment of the inherent
relation between operators such as equ(\ref{inherent}). Furthermore
such multi-hidden variable density matrix has one very important
implication, that according to equ(\ref{hvtcondition}), all states
(events) corresponding to arbitrary directions should all be
exclusive events. This implies
$\sigma_{\vec{r}_{1}}\sigma_{\vec{r}_{2}}=0$ and our CPT density
matrix has to be
\begin{equation}\label{rho_exclusive}
\rho^{hvt} =
\frac{1}{\mathcal{N}}\sum_{\vec{r}}\left[p_{\uparrow}\left(\vec{r}\right)\left|\uparrow\right>_{\vec{r}}\left<\uparrow\right|_{\vec{r}}
+
p_{\downarrow}\left(\vec{r}\right)\left|\downarrow\right>_{\vec{r}}\left<\downarrow\right|_{\vec{r}}\right]
\end{equation}
where
\begin{equation}
p_{\uparrow}\left(\vec{r}\right) = \frac{1+r_{x}}{2},
p_{\downarrow}\left(\vec{r}\right) = \frac{1-r_{x}}{2}.
\end{equation}

There is a technical problem and another non-trivial
conceptual problem with the above PDF. The technical problem is the
value of $\mathcal{N}$. Since we need to keep
$tr\left(\rho^{hvt}\right)=1$ and there is infinite number of
directions, $\mathcal{N}$ will be infinity if
$tr\left(\rho^{hvt}\right)$ is simply,
\begin{equation}\label{trace1}
tr\left(\rho^{hvt}\right)
=\sum_{\vec{r}}\left[\left<\uparrow\right|_{\vec{r}}\rho^{hvt}\left|\uparrow\right>_{\vec{r}}
+
\left<\downarrow\right|_{\vec{r}}\rho^{hvt}\left|\downarrow\right>_{\vec{r}}\right].
\end{equation}
One way to define a ``proper'' $tr\left(\rho^{hvt}\right)$ to avoid
such divergence is to decompose $\vec{r} =
\left(\sin{\theta}\cos{\phi},\sin{\theta}\sin{\phi},
\cos{\theta}\right)$ and treat
\begin{equation}\label{trace2}
tr\left(\rho^{hvt}\right) =\int d\theta d\phi
\sin{\theta}\left[\left<\uparrow\right|_{\vec{r}}\rho^{hvt}\left|\uparrow\right>_{\vec{r}}
+
\left<\downarrow\right|_{\vec{r}}\rho^{hvt}\left|\downarrow\right>_{\vec{r}}\right].
\end{equation}
In this case, $\mathcal{N}=4\pi$. This introduces additional
relative probability between states corresponding to different
$\vec{r}$. This may not be a proper definition, however, it is
still possible to solve this technical question of divergent
normalization constant by some other ways. If only relative
probability of a given direction $\vec{r}$ is concerned in real
measurements, this problem does not affect the outcomes at all.

The other problem is rather serious. That is due to the full
exclusiveness between all events, the meaning of a measurement changes.
``Measuring $\sigma_{\vec{r}}$'' for a specifically given $\vec{r}$
is no longer a pure elementary event but a compound event. A
pure elementary event instead would be ``measuring $\sigma$'', with
no specific direction given. The result of such a measurement
will be one direction, which got randomly picked up during the
measurement process, and an $up$- or $down$- state, would be recorded
correspondingly with the right probability. In this way, there is no
guarantee that the randomly picked-up direction will be the desired
direction of an observer.

A classical dice would be a good example of a classical
probability distribution based on all exclusive events. From a perfect
$6$-face dice, we wish to only measure the relative probability
between face $1$ and face $2$. We
could still get all $6$ numbers, but we discard all the other four if they
turn out to be the outcomes of our measurement. Therefore,
effectively we will find out the state of the dice within the
subspace is,
\begin{equation}
\rho^{c}=\frac{1}{2}\left(\left|1\right>\left<1\right| +
\left|2\right>\left<2\right|\right).
\end{equation}
Similarly measurement of $\sigma$ on our $\rho^{hvt}$ will
be one of all of the exclusive events, and during our analysis of the results, we
can discard all irrelevant events. In real quantum
measurements, however, we never find such irrelevant and redundant outcomes.
If we measure $\sigma_{x}$, according to the all
exclusive nature, in a classical measurement of the above state,
sometimes our apparatus detects nothing and sometimes it detects the
right state -- $up$. However, in a real quantum measurement,
assuming no further experimental accuracy limits, such
detecting-nothing events never happen. What we get is only the $up$- or
$down$- state of a given direction. This shows that in fact the
above state based on exclusiveness is not the desired state. Or if
it is then this is only possible if the system somehow knows
the intention of the observer during the process.

This ``contextual'' relation between system and observer is
unexpected, however, some physicists may still be willing to accept
such a theory since it is a problem about interpretation of a
theory not about any predictions from the theory.

Now we will try to make this HVT compatible with QS-III and QS-IV, those
facts about repeat measurement. CPT-III tells us that if all events
are exclusive, then after a measurement, for example along the $x$
direction, given the $up$ state is recorded, its state is simply
$\left|\uparrow\right>_{x}\left<\uparrow\right|_{x}$. A repeat
measurement along the $x$ direction results in event $up$ again.
This is the expected result stated in QS-III. However, if the repeat
measurement is along the $z$ direction, for state
$\left|\uparrow\right>_{x}\left<\uparrow\right|_{x}$, the
exclusiveness tells us, there is not any such events as measurement
along $z$ direction. So we would again get a detecting-nothing
event. This conflicts with QS-IV. Furthermore, the
initial state is the $up$ state along the $x$ direction, therefore after a
measurement along the $x$ direction, nothing changes. If
CPT-III holds, we see the state before and after the measurement is
respectively,
\begin{equation}\label{beforemeasure}
\rho^{hvt}_{before} =
\sum_{\vec{r}}\left[p_{\uparrow}\left(\vec{r}\right)\left|\uparrow\right>_{\vec{r}}\left<\uparrow\right|_{\vec{r}}
+
p_{\downarrow}\left(\vec{r}\right)\left|\downarrow\right>_{\vec{r}}\left<\downarrow\right|_{\vec{r}}\right]
\end{equation}
and
\begin{equation}\label{aftermeasure}
\rho^{hvt}_{after} =
\left|\uparrow\right>_{x}\left<\uparrow\right|_{x},
\end{equation}
where
\begin{equation}
p_{\uparrow}\left(\vec{r}\right) = \frac{1+r_{x}}{2},
p_{\downarrow}\left(\vec{r}\right) = \frac{1-r_{x}}{2}.
\end{equation}
Equ\eqref{beforemeasure} and equ\eqref{aftermeasure} are obviously
different. The state stays the same before and after the measurement,
however, we find their expressions are different. This means
CPT-III is wrong. A state after it was revealed in a
measurement is not the state corresponding to the measurement
result. We will have to also sacrifice CPT-III after abandoning
equ\eqref{inherent}.
\begin{list}{CPT-\Roman{Lcount}$^{'}$}
{\usecounter{Lcount}
    \setlength{\rightmargin}{\leftmargin}}
\setcounter{Lcount}{2}
  \item After a measurement, the object stays at the state which guarantees
  a repeat measurement in accordance with QS-III and
  QS-IV. For example, for spin-$\frac{1}{2}$ after a measurement on $\sigma_{\vec{r}_{0}}$ and
  a $up$-state being recorded, the state is,
\begin{equation}\label{CPTIII'}
\rho^{hvt}_{after} =
\sum_{\vec{r}}\left[\frac{1+\vec{r}\cdot\vec{r_{0}}}{2}\left|\uparrow\right>_{\vec{r}}\left<\uparrow\right|_{\vec{r}}
+
\frac{1-\vec{r}\cdot\vec{r_{0}}}{2}\left|\downarrow\right>_{\vec{r}}\left<\downarrow\right|_{\vec{r}}\right].
\end{equation}
If a $down$-state is recorded after measurement of
$\sigma_{\vec{r}_{0}}$, we can simply replace $\vec{r}_{0}$
with $-\vec{r}_{0}$ in equ\eqref{CPTIII'}.
\end{list}
This CPT-III$^{'}$ is not easily understood.

Furthermore, this is not a consistent theory. We already know that
an $up$-state along the $x$ direction, before and after measurements of
$\sigma_{x}$, is equ\eqref{beforemeasure}. Given this state if we
want to calculate the probability of observing the $up$-state along the $x$ direction,
we will do
\begin{equation}
\left\{
\begin{aligned}
p_{x_{up}} =
\left<\uparrow\right|_{x}\rho^{hvt}_{before}\left|\uparrow\right>_{x}
= \frac{1}{\mathcal{N}},
\\ p_{x_{down}} =
\left<\downarrow\right|_{x}\rho^{hvt}_{before}\left|\downarrow\right>_{x}
= 0.
\end{aligned}
\right.
\end{equation}
This gives the correct answer that the relative probability between
$up$- and $down$-state is $1$. But notice that we times
$\left<\uparrow\right|_{x}$ from the left and
$\left|\uparrow\right>_{x}$ the right to a density matrix to get the
probability of $p_{x_{up}}$. In doing so we assume that vector
$\left|\uparrow\right>_{x}$ stands for the event of an $up$-state along the $x$
direction, but it is different from equ\eqref{beforemeasure}, which
is the expression standing for the event of an $up$-state along the $x$
direction as we pointed out before. We have now two different
expressions for the same state in a theory.

Therefore we conclude that the first rescue of HVT, based on the
assumption of the exclusiveness among all
$\left\{\sigma_{\vec{r}}\right\}$, failed to achieve a consistent
theory satisfying simultaneously QS-I, QS-II, QS-III and QS-IV. To
do so we will not have inherent relation between operators as in
equ\eqref{inherent}, we will have to put a twist on CPT-III and allow
``contextual'' communication between object and observer. Even after
all these, we would not be able to get a self-consistent theory. We
will now try out another more plausible construction of CPT for QS,
based on independence of all pure elementary events.

\subsection{CPT based on independence of pure elementary events}
Although the idea of exclusive events fails, in fact, there is
another way to save the idea of HVT, being that $\lambda_{z}$ and
$\lambda_{x}$ are independent events, so that a HVT density matrix
could be,
\begin{equation}
\rho^{hvt} =
\sum_{\left\{\vec{\lambda}\right\}}\rho\left(\vec{\lambda}\right)\left|\dots,
x_{\vec{r}}\left(\lambda_{\vec{r}}\right), \dots\right>\left<\dots,
x_{\vec{r}}\left(\lambda_{\vec{r}}\right), \dots\right|,
\label{bohmhvt}
\end{equation}
where $\lambda_{\vec{r}}$ is a random variable for direction
$\vec{r}$ and $x_{\vec{r}}\left(\lambda_{\vec{r}}\right) = \uparrow,
\downarrow$. Notation $\vec{\lambda}$ refers to an infinite
dimensional vector $\left(\lambda_{x}, \cdots, \lambda_{y}, \cdots,
\lambda_{z}, \cdots \right)$. Under this independent event
assumption, measurement on every direction is done independently.
This is only possible if all $\sigma_{\lambda_{\vec{r}}}$,
operators corresponding to all directions $\lambda_{\vec{r}}$ are
commutative and every operator could be treated independently.

HVT requires again abandoning inherent relation as in
equ\eqref{inherent} and non-commutative relation between quantum
operators. A valid multiplication between operators is the direct
product, $\sigma_{\vec{r}_{1}}\otimes\sigma_{\vec{r}_{2}}$. A common
basis is $\left|\left(\uparrow_{x} \mbox{ or }\downarrow_{x}\right),
\cdots, \left(\uparrow_{y} \mbox{ or }\downarrow_{y}\right),
\cdots,\left(\uparrow_{z} \mbox{ or }\downarrow_{z}\right), \cdots
\right>$. We have an infinite number of hidden variables to
represent all directions of measurement. Principally, by choosing
appropriate $\rho\left(\vec{\lambda}\right)$ one can always fulfill
QS-I and II. For example, the following scheme gives the correct
results on measurement of $\sigma_{r}$. For direction $\vec{r} =
\left(r_{x}, r_{y}, r_{z}\right)$, we choose
$\lambda_{\vec{r}}\in\left\{-\frac{1}{2},\frac{1}{2}\right\}$, a
two-value discrete random variable as the hidden variable. Then we
require the partial trace except direction $\vec{r}$ of $\rho^{hvt}$
gives,
\begin{equation}
\rho^{hvt}_{\vec{r}}\triangleq tr^{-\vec{r}}\left(\rho^{hvt}\right)
=
\frac{1+r_{x}}{2}\left|\uparrow\right>_{\vec{r}}\left<\uparrow\right|_{\vec{r}}
+
\frac{1-r_{x}}{2}\left|\downarrow\right>_{\vec{r}}\left<\downarrow\right|_{\vec{r}},
\label{example}
\end{equation}
for example, by requiring
\begin{equation}\label{product-state}
\rho^{hvt}=\Pi_{\vec{r}}\otimes\rho^{hvt}_{\vec{r}}.
\end{equation}
Notice the product state in equ\eqref{product-state} is just one
example, not necessary required, while equ\eqref{example} is a
strict requirement. There are many more density matrices in the form
of equ\eqref{bohmhvt} and satisfying equ\eqref{example}.
Independence of pure elementary events does not lead to independent
product states.

One can check this satisfies QS-I and II for measurement on
an arbitrary $\vec{r}$ direction. The outcomes could be $\uparrow$ or
$\downarrow$ with probability of $\frac{1+r_{x}}{2}$ and
$\frac{1-r_{x}}{2}$ respectively. Furthermore, this HVT does not require
contextuality between object and observer. After the partial trace
only the desired direction will survive. The partial trace is a
standard procedure for independent random variable. The above
explicitly constructed density matrix gives the correct results for
measurement on any directions. We see that our HVT is at least as
valid as Bell's HVT on a spin $\frac{1}{2}$ object\cite{bellrmp} as
they both respect QS-I and QS-II. It is less controversial than
our former exclusive-event HVT.

Will this HVT realize QS-III and QS-IV? The answer is ``yes'' for
QS-III. According to CPT-III, after measurement, the system stays at
the state observed for the observable and all the others remain at the
same states. For example, when we measure $\sigma_z$ with the
outcome being $up$, the state after measurement is
\begin{equation}
\rho^{hvt}_{after} =\frac{
\left|\uparrow\right>_{z}\left<\uparrow\right|_{z}\otimes\left<\uparrow\right|_{z}\rho^{hvt}_{before}\left|\uparrow\right>_{z}}{tr^{-z}\left(\left<\uparrow\right|_{z}\rho^{hvt}_{before}\left|\uparrow\right>_{z}\right)}.
\label{finalstate}
\end{equation}
If measured on $\sigma_{z}$ again the outcome is still $up$. What if the second measurement is on a different
direction, say $\sigma_{x}$? We have,
\begin{align}
p_{{z_{up}},{x_{down}}} & = & tr\left(
\left|\downarrow\right>_{x}\left<\downarrow\right|_{x}\rho^{hvt}_{after}\right)
\notag
\\ & = & \frac{tr^{-x,-z}\left(
\left<\uparrow\right|_{z}\left<\downarrow\right|_{x}\rho^{hvt}_{before}\left|\downarrow\right>_{x}\left|\uparrow\right>_{z}\right)}{tr^{-z}\left(\left<\uparrow\right|_{z}\rho^{hvt}_{before}\left|\uparrow\right>_{z}\right)}
\notag
\\ & = & 0,
\end{align}
where we make use of
$\left<\downarrow\right|_{x}\rho^{hvt}_{before}\left|\downarrow\right>_{x}=0$,
the fact that the state is initially $x$ direction $up$. However,
this number is expected to be $\frac{1}{2}$. This shows our HVT does
not respect QS-IV if CPT-III holds. Thus we need to modify
CPT-III to the following,
\begin{list}{CPT-\Roman{Lcount}$^{''}$}
{\usecounter{Lcount}
    \setlength{\rightmargin}{\leftmargin}}
    \setcounter{Lcount}{2}
  \item  After a measurement, the object stays at the state which guarantees
  a repeat measurement gives the right result stated in QS-III and
  QS-IV. For example, for spin-$\frac{1}{2}$ after a measurement on $\sigma_{\vec{r}_{0}}$ and
  a $up$-state is recorded, it stays at, $\rho^{hvt}$ satisfying $\rho^{hvt}_{\vec{r}}=
tr^{-\vec{r}}\left(\rho^{hvt}\right)$,
\begin{equation} \rho^{hvt}_{\vec{r}}=
\frac{1+\vec{r}\cdot\vec{r}_{0}}{2}\left|\uparrow\right>_{\vec{r}}\left<\uparrow\right|_{\vec{r}}
+
\frac{1-\vec{r}\cdot\vec{r}_{0}}{2}\left|\downarrow\right>_{\vec{r}}\left<\downarrow\right|_{\vec{r}},
\label{example''}
\end{equation}
If a $down$-state is recorded after measurement of
$\sigma_{\vec{r}_{0}}$, we can simply replace $\vec{r}_{0}$
with $-\vec{r}_{0}$ in equ\eqref{example''}.
\end{list}

This version of CPT-III has the same inconsistency
with the last exclusive-event HVT. Given a $x$ direction $up$-state, represented by equ\eqref{example},
if we want to calculate the probability of $x$ direction $up$-state,
we do
\begin{equation}
p_{x_{up}}=tr\left(\left|\uparrow\right>_{x}\left<\uparrow\right|_{x}\rho^{hvt}\right)=1.
\end{equation}
This gives us the correct result, however, with the assumption that
$\left|\uparrow\right>_{x}\left<\uparrow\right|_{x}$ refers to the
$x$ direction $up$-state, which is not the state which really means
the $x$ direction $up$-state as in equ\eqref{example}.

\subsection{Final HVT: equivalent class on the set of density matrices}
We have noticed that the density matrix in product form as in
equ\eqref{product-state} is just a special case of
equ\eqref{bohmhvt}. Due to equ\eqref{example}, all
qualified density matrices give correct results on measurement of
$\sigma_{\vec{r}}$ along an arbitrary direction $\vec{r}$. One may
tell the difference among them if measurements along different
directions are performed simultaneously. Either because reality
forbids us to do so, or because we do not have the technology yet,
we are not able to perform such measurements. Before the possibility
of those simultaneous measurements can be resolved, we do not know
principally which specific one out of the set of general form is the
right description of QS.

For now, we will focus on measurement along one direction.
For this case, our density matrix based description of quantum system is
redundant. An equivalent class over the whole set of density
matrices in the general form of equ\eqref{bohmhvt} can be defined as following: Two
density matrices $\rho^{a}, \rho^{b}$ are regarded as equivalent if
and only if they lead to the same reduced density matrix,
\begin{equation}
tr^{-\vec{r}}\left(\rho^{a}\right) =
tr^{-\vec{r}}\left(\rho^{b}\right), \forall \vec{r}\in
\mathbb{R}^{3}.
\end{equation}
Our state is represented by the those equivalent class as required by QS-V.

For example, the first preparation gives us state,
\begin{equation}
\rho^{I}=\frac{1}{4}\rho^{11} + \frac{3}{4}\rho^{12},
\end{equation}
$\rho^{11}$ is a product state of $\rho^{11}_{\vec{r}}$,
\begin{equation}
\rho^{11}_{\vec{r}}=\frac{1+r_{z}}{2}\left|\uparrow\right>_{\vec{r}}\left<\uparrow\right|_{\vec{r}}
+
\frac{1-r_{z}}{2}\left|\downarrow\right>_{\vec{r}}\left<\downarrow\right|_{\vec{r}}.
\end{equation}
$\rho^{12}$ is a product state of $\rho^{12}_{\vec{r}}$,
\begin{equation}
\rho^{12}_{\vec{r}}=\frac{1-r_{z}}{2}\left|\uparrow\right>_{\vec{r}}\left<\uparrow\right|_{\vec{r}}
+
\frac{1+r_{z}}{2}\left|\downarrow\right>_{\vec{r}}\left<\downarrow\right|_{\vec{r}}.
\end{equation}
The second preparation gives us state,
\begin{equation}
\rho^{II}=\frac{1}{2}\rho^{21} + \frac{1}{2}\rho^{22},
\end{equation}
$\rho^{21}$ is a product state of $\rho^{21}_{\vec{r}}$,
\begin{equation}
\rho^{21}_{\vec{r}}=\frac{1-r_{y}\frac{\sqrt{3}}{2} -
r_{z}\frac{1}{2}}{2}\left|\uparrow\right>_{\vec{r}}\left<\uparrow\right|_{\vec{r}}
+ \frac{1+r_{y}\frac{\sqrt{3}}{2} +
r_{z}\frac{1}{2}}{2}\left|\downarrow\right>_{\vec{r}}\left<\downarrow\right|_{\vec{r}}.
\end{equation}
$\rho^{22}$ is a product state of $\rho^{22}_{\vec{r}}$,
\begin{equation}
\rho^{22}_{\vec{r}}=\frac{1+r_{y}\frac{\sqrt{3}}{2} -
r_{z}\frac{1}{2}}{2}\left|\uparrow\right>_{\vec{r}}\left<\uparrow\right|_{\vec{r}}
- \frac{1+r_{y}\frac{\sqrt{3}}{2} +
r_{z}\frac{1}{2}}{2}\left|\downarrow\right>_{\vec{r}}\left<\downarrow\right|_{\vec{r}}.
\end{equation}
In fact, we can also write down a product form density matrix
according to equ\eqref{rho_QSV}. $\rho^{III}$ is a product state of
$\rho^{III}_{\vec{r}}$,
\begin{equation}
\rho^{III}_{\vec{r}}=\frac{2-r_{z}}{4}\left|\uparrow\right>_{\vec{r}}\left<\uparrow\right|_{\vec{r}}
+
\frac{2+r_{z}}{4}\left|\downarrow\right>_{\vec{r}}\left<\downarrow\right|_{\vec{r}}
\end{equation}
It is straightforward to check that
\begin{equation}
tr^{-\vec{r}}\left(\rho^{I}\right)
=tr^{-\vec{r}}\left(\rho^{II}\right)
=tr^{-\vec{r}}\left(\rho^{III}\right).
\end{equation}
But
\begin{align}
\rho^{I}\neq\rho^{III}\neq\rho^{II}.
\end{align}
Without the equivalent class, they are different density matrices.
Then our HVT does not respect QS-V. With it, they are regarded as
the same so that QS-V is satisfied.

If in the future, we would be able to measure $\sigma$ along
several directions simultaneously, it would force us to pick up one
specific form out of the whole set satisfying equ\eqref{example} and
equ\eqref{bohmhvt} and to discard the above
equivalent class. For now, this is the final HVT we can propose as far
as we require it to respect all five QS facts.

\subsection{Test of all above HVTs against measurements}
Besides our EHVT and IHVT, let us also check Bell's HVT and Bohm's HVT against our five QS facts.
Imagine we are given one of the five states below and a measurement device as
will be explained. Then we are asked to find out which one
is the real state of the given object.
\begin{enumerate}
\renewcommand{\labelenumi}{\Alph{enumi}}
\item A quantum spin-$\frac{1}{2}$ at
$\rho_{0}=\left|\uparrow\right>_{x}\left<\uparrow\right|_{x}$.
\item A classical two-face dice at state $\rho_{0}=\frac{1}{2}\left|\uparrow\right>_{z}\left<\uparrow\right|_{z} +
\frac{1}{2}\left|\downarrow\right>_{z}\left<\downarrow\right|_{z}$.
\item A classical vector pointing to arbitrary directions with
probability $\rho_{0}=\frac{1}{4\pi}\iint d\theta
d\phi\sin{\theta}\frac{1+\sin{\theta}\cos{\phi}}{2}\left|\uparrow\right>_{\vec{r}}\left<\uparrow\right|_{\vec{r}}$.
If we rewrite state A in a spin coherent basis, we will get the same
distribution. The only difference is that here in treating it like a
classical object, we further assume the basis is orthogonal. It is
a state in the form of an exclusive-event HVT (EHVT).
\item A classical object at state $\rho_{0}=\Pi_{\vec{r}\in \mathbb{D}}\otimes
\rho_{0}^{\vec{r}}$, where
$\rho_{0}^{\vec{r}}=\frac{1+\sin{\theta}\cos{\phi}}{2}\left|\uparrow\right>_{\vec{r}}\left<\uparrow\right|_{\vec{r}}
+
\frac{1-\sin{\theta}\cos{\phi}}{2}\left|\downarrow\right>_{\vec{r}}\left<\downarrow\right|_{\vec{r}}$.
Here $D=\left([0,\frac{\pi}{2})\otimes[0,2\pi)\right)\cup
\left(\left\{\frac{\pi}{2}\right\}\otimes[0,\pi)\right)$, which
denotes half of all direction vector $\vec{r}$. This is a state in
the form of an independent-event-equivalent-class HVT (IHVT).
\item Bell's hidden variable theory of spin-$\frac{1}{2}$
object\cite{bellrmp}. Hidden variable $\lambda\in\left[-\frac{1}{2},
\frac{1}{2}\right]$ uniformly distributed. Given a specific
$\lambda$, measurement on Pauli matrix
$\vec{\beta}\cdot\vec{\sigma}$ on direction $\vec{\beta}$ yields,
$sign\left({\lambda+\frac{1}{2}\beta_{x}}\right)sign\left({X}\right)$,
where $X=\beta_{x}$ if $\beta_{x}\neq0$, $X=\beta_{y}$ if
$\beta_{y}\neq0, \beta_{x}=0$ and $X=\beta_{z}$ if $\beta_{z}\neq0,
\beta_{x}=0, \beta_{y}=0$. Here we changed the expression
accordingly to represent the $x$ direction $up$ state.
\end{enumerate}
The measurement device has an indicator showing a positive/negative value if the
object is along the same/opposite direction. One can control
the direction of the device. When its direction is not parallel or
opposite to the object's direction, it will not be activated. Assume
this device is sharp so that it will not respond to even a slight
mis-matching. The device works on both classical and quantum
systems.

Define the activation ratio $Q$ as the ratio between times when the
device is activated out of the total times the device is used, and
define the $up$-state probability $P$ as the ratio between the numern of positive values out
of the times when the device is activated. We want to check if the
above five states give us different values of $Q$ and $P$ during
measurements. First, assume the device is along the $z$ direction.
\begin{table}[!h]
\tabcolsep 0pt \caption{Values of $Q$ and $P$ with device along the $z$
direction} \vspace*{-24pt}
\begin{center}
\def\temptablewidth{0.5\textwidth}
{\rule{\temptablewidth}{1pt}}
\begin{tabular*}{\temptablewidth}{@{\extracolsep{\fill}}|c|c|c|c|c|c|}
& A & B &C &D &E\\  \hline $Q$ & $1$ & $1$ & $\ll1$\cite{footnote} & $1$ & $1$\\ \hline $P$ & $0.5$ & $0.5$ & $0.5$ & $0.5$ & $0.5$\\
\end{tabular*}
{\rule{\temptablewidth}{1pt}}
\end{center}
\end{table}
We see from Table I that from the values of $Q$, state $C$ is different from state
$A$.

Next we adjust the device to the $x$ direction.
\begin{table}[!h]
\tabcolsep 0pt \caption{Values of $Q$ and $P$ with device along the $x$
direction} \vspace*{-24pt}
\begin{center}
\def\temptablewidth{0.5\textwidth}
{\rule{\temptablewidth}{1pt}}
\begin{tabular*}{\temptablewidth}{@{\extracolsep{\fill}}|c|c|c|c|c|c|}
& A & B &C &D & E\\  \hline $Q$ & $1$ & $0$ & $\ll1$ & $1$ & $1$\\ \hline $P$ & $1$ & NA & $1$ & $1$ & $1$\\
\end{tabular*}
{\rule{\temptablewidth}{1pt}}
\end{center}
\end{table}
We see from Table II that from the values of $Q$ state $B$ is different from state
$A$. However, those measurements do not differentiate state $A$, $D$
and $E$. The fact those two states $D$ and $E$ both respects QS-I
and QS-II, makes them very good counterexamples of Von Neumann's
proof of impossibility of HVT\cite{von}. This is exactly made
possible by that operators in those two theories do not obey
\eqref{inherent} the linear relation between operators even when
operators' averages have those linear relation. Such relation
between operators is too restrictively assumed in Von Neumann's
proof and leads to impossibility\cite{von, bohmrmp}.

In order to differentiate state $A$, $D$ and $E$, we have to perform
repeat measurement, say first along the $z$ direction and then along the $x$
direction. In dealing with repeat measurement, we need some rules to
determine the object's state right after the first measurement. Here we
first assume both CPT-III and QM-III hold. In the following table,
we list only values of $Q$ and $P$ after the second measurement.
\begin{table}[!h]
\tabcolsep 0pt \caption{Values of $Q$ and $P$ during the second
measurement in a repeat measurement with device along the $z$ and then
the $x$ direction, assuming both CPT-III and QM-III hold}
\vspace*{-24pt}
\begin{center}
\def\temptablewidth{0.5\textwidth}
{\rule{\temptablewidth}{1pt}}
\begin{tabular*}{\temptablewidth}{@{\extracolsep{\fill}}|c|c|c|c|c|c|}
& A & B &C &D & E\\
 \hline $Q_{2}$ & $1$ & $0$ & $0$ & $1$ & $1$\\ \hline $P_{2}$ & $0.5$ & NA & NA & $1$ & $1$\\
\end{tabular*}
{\rule{\temptablewidth}{1pt}}
\end{center}
\end{table}
From Table III the values of $P$ there we find that state $D$ and $E$ are
different from state $A$. That is we can distinguish a quantum state
with Bell's HVT and our IHVT state by measurements if CPT-III/QM-III
holds. As for state $D$, this can be seen from,
\begin{equation}
\rho_{1} =\frac{
\left|\uparrow\right>_{z}\left<\uparrow\right|_{z}\otimes\left<\uparrow\right|_{z}\rho_{0}\left|\uparrow\right>_{z}}
{tr^{-z}\left(\left<\uparrow\right|_{z}\rho_{0}\left|\uparrow\right>_{z}\right)},
\end{equation}
and
\begin{equation}
P_{2} = \frac{
tr^{-x,-z}\left(\left<\uparrow\right|_{z}\left<\uparrow\right|_{x}\rho_{0}\left|\uparrow\right>_{x}\left|\uparrow\right>_{z}\right)}
{tr^{-z}\left(\left<\uparrow\right|_{z}\rho_{0}\left|\uparrow\right>_{z}\right)}=1.
\end{equation}
As for state $E$, let's assume $\lambda=\lambda^{*}$ after the first
measurement, then for the second measurement one will get,
\begin{align}
sign\left({\lambda^{*}+\frac{1}{2}\beta_{x}}\right)sign\left(\beta_{x}\right)
= sign\left({\lambda^{*}+\frac{1}{2}}\right)=1, \forall \lambda^{*}.
\end{align}
If we are allowed to relax CPT-III then it is always possible to
adjust $\rho_{1}$ for state $D$ and adjust the proposed measurement
result for state $E$ after the first measurement to make
$P_{2}=0.5$. We have done so for state $D$ in CPT-III$^{''}$. And
here we can adjust state $E$ to satisfy the requirement. That is if
we get the $up$/$down$-state in the first measurement, for arbitrary
second measurement of $\vec{\beta}\cdot\vec{\sigma}$
\begin{align}
sign\left({\lambda\pm\frac{1}{2}\beta_{z}}\right)sign\left({X}\right),
\end{align}
where
\begin{equation}
X=\left\{ \begin{aligned}
         \beta_{z} & \mbox{ if } \beta_{z}\neq0\\
         \beta_{x} & \mbox{ if } \beta_{z}=0, \beta_{x}\neq0\\
         \beta_{y} & \mbox{ if } \beta_{z}=0,\beta_{x}=0, \beta_{y}\neq0
        \end{aligned} \right..
\end{equation}
In that case, state $D$ and $E$ are indistinguishable from state $A$
under all measurements, while state $D$ and $E$ are classical states
and state $A$ is a quantum state, as we see in Table IV.
\begin{table}[!h]
\tabcolsep 0pt \caption{Values of $Q$ and $P$ during the second
measurement in a repeat measurement with device along $z$ and then
$x$ direction, with CPT-III adjusted accordingly} \vspace*{-24pt}
\begin{center}
\def\temptablewidth{0.5\textwidth}
{\rule{\temptablewidth}{1pt}}
\begin{tabular*}{\temptablewidth}{@{\extracolsep{\fill}}|c|c|c|c|c|c|}
& A & B &C &D &E\\
 \hline $Q_{2}$ & $1$ & $0$ & $\ll1$ & $1$ & $1$\\ \hline $P_{2}$ & $0.5$ & NA & $0.5$ & $0.5$ & $0.5$\\
\end{tabular*}
{\rule{\temptablewidth}{1pt}}
\end{center}
\end{table}

From above comparison, we see that when repeat measurement is taken
into consideration and CPT-III holds, none of all five theories respects all five QS
facts. Only when we relax CPT-III, both Bell's
HVT and our IHVT provide alternative theory for quantum systems.
Unlike Bell's HVT theory, in our IHVT state, we have explicitly
written down the state in a density matrix form, so it can be
generalized for any objects not only spin-$\frac{1}{2}$ particles.
In this sense this work can be seen as a development of Bell's HVT.
Another thing we would like to point out is the relation between our
IHVT and Bohm's HVT\cite{bohmrmp}: in the following sense, our IHVT
provides exactly the explicit form of a state of Bohm's HVT.

Originally Bohm's HVT gave only a classical HVT based interpretation
of measurement process on a single direction. Since we are free to
choose an arbitrary direction, we need to generalize the theory a
little bit. Basically it says during measurement process of a
specific direction $\vec{r}$, system evolves according to the
following equation system,
\begin{align}\label{bohmmeasure}
\left\{ \begin{array}{c} \frac{dJ^{1}_{\vec{r}}}{dt} =
2\gamma\left(R^{1}-R^{2}\right)J^{1}_{\vec{r}}J^{2}_{\vec{r}}
\\
\frac{dJ^{2}_{\vec{r}}}{dt} =
2\gamma\left(R^{2}-R^{1}\right)J^{2}_{\vec{r}}J^{1}_{\vec{r}}
\end{array}\right.,
\end{align}
where
$R^{i}=\frac{\left|J^{i}_{\vec{r}}\right|^{2}}{\left|\xi^{i}_{\vec{r}}\right|^{2}}$
and $\xi^{i}_{\vec{r}}$ are those hidden variables. Instead of
quantum wavefunction $\psi$ here we take $J^{i}_{\vec{r}}$ as our
fundamental variables since only
$J^{1}_{\vec{r}}=\left|\left<\uparrow_{\vec{r}}\left|\right.\psi\right>\right|^{2}$
and
$J^{2}_{\vec{r}}=\left|\left<\downarrow_{\vec{r}}\left|\right.\psi\right>\right|^{2}$
is used in those equations. Then if we only focus on state
representing this direction only, it can be written down as
\begin{align}\label{bohmstate}
\rho_{\vec{r}} =
J^{1}_{\vec{r}}\left|\uparrow_{\vec{r}}\left>\right<\uparrow_{\vec{r}}\right|
+
J^{2}_{\vec{r}}\left|\downarrow_{\vec{r}}\left>\right<\downarrow_{\vec{r}}\right|.
\end{align}
From this point of view, \eqref{bohmmeasure} provides an explanation
of the process that the above state in \eqref{bohmstate} turns into
$\left|\uparrow_{\vec{r}}\left>\right<\uparrow_{\vec{r}}\right|$ or
$\left|\downarrow_{\vec{r}}\left>\right<\downarrow_{\vec{r}}\right|$
at probability respectively $J^{1}_{\vec{r}}$ or $J^{2}_{\vec{r}}$.
Now let's consider a separate measurement along another direction
$\vec{r}^{'}$. One possible way is to start from the quantum
wavefunction $\psi$ again to calculate $J^{i}_{\vec{r}^{'}}$ and
redo above procedure. This is actually not so bad but in this way
this theory never gets rid of the quantum wavefunction. This makes
Bohm's HVT only an alternative theory of quantum measurements but
not a coherent theory covering both evolutions and measurements.
There is however another way to recover the right prediction of
measurement on $\vec{r}^{'}$ and it gets ride of quantum
wavefunction totally. That is to assume that the HVT state is in
fact,
\begin{align}
\rho = \Pi_{\vec{r}} \otimes \rho_{\vec{r}},
\end{align}
while $\rho_{\vec{r}}$ is given by \eqref{bohmstate} for a specific
direction $\vec{r}$ with proper predefined $J^{i}_{\vec{r}}$. We see
that this is exactly state $D$, our IHVT state. We have shown that
this state agrees with quantum mechanical state $A$ on everything
with CPT-III replaced by CPT-III$^{''}$.

However, this IHVT is far from a standard CPT. To summarize, IHVT
satisfies QS-I and QS-II easily but CPT-III needs to be modified to
make it satisfy QS-III and QS-IV. IHVT does not require
contextuality between object and observer as EHVT does. But both
suffer from the same inconsistency problem: two different
expressions are used to represent the same state for two different
purposes. Furthermore, both discard inherent relation among
operators as in equ\eqref{inherent} by treating operators
independently or exclusively. We find that all of the above has made
HVT less understandable than the usual QM, which has none of above
problems and respects all five QS facts. Therefore, we would like to
conclude that we have ruled out HVT just from theoretical
consideration and just by considering a spin-$\frac{1}{2}$ object.

If one is still willing to pay all the prices mentioned above, then we are also
willing to go a little further to show that this IHVT conceals
something else which one may not want in a theory of physics. We will
apply this theory onto the description of singlet state, the
entangled state used in the discussion of Bell's inequality.

\section{CPT for two-spin system}
\label{sechidden2}
The quantum density matrix form of a singlet
state is
\begin{equation}
\rho^{q} = \frac{1}{2}\left(\left|\uparrow\downarrow\right>
 - \left|\downarrow\uparrow\right>\right)\left(\left<\uparrow\downarrow\right|
 - \left<\downarrow\uparrow\right|\right),
\end{equation}
where $\left|\uparrow\downarrow\right>$ can be regarded as
eigenstates on an arbitrary direction. The correlated quantum
measurement of the $\vec{r}_{1}$-direction on spin $1$ and $\vec{r}_{2}$
on spin $2$ gives
\begin{equation}
\left<\sigma_{\vec{r}_{1}}\sigma_{\vec{r}_{1}}\right> =
-\vec{r}_{1}\cdot\vec{r}_{2} \text{ and }
\sigma_{\vec{r}_{1}}\sigma_{\vec{r}_{2}} = \pm1. \label{correlate}
\end{equation}
A measurement on a single spin along any direction gives
\begin{equation}
\left<\sigma_{\vec{r}}\right> = 0 \text{ and } \sigma_{\vec{r}} =
\pm1. \label{single}
\end{equation}
A successful HVT theory should give the two above results. Besides,
for a repeat measurement, HVT should also give the correct results
depending on the outcome from the first measurement. Although Bell's
inequality has generally proved that through local classical theory
it is impossible to achieve this, here, we will construct one such
state, in the form of a classical density matrix, that does in fact
achieve this. We will the find out the cost of such a theory.

To denote a state in IHVT, one example of an equivalent class is
used to represent the whole class. We should check if the following
state respects all the QS facts. A reduced density matrix for two spins on
$\hat{z}$ and $\hat{r}$ is
\begin{equation}
\begin{array}{lll}
\rho_{\vec{r}_{1},\vec{r}_{2}} & = &
\frac{1-\vec{r}_{1}\cdot\vec{r}_{2}}{4}\left(\left|\uparrow_{\vec{r}_{1}}\uparrow_{\vec{r}_{2}}\right>
\left<\uparrow_{\vec{r}_{1}}\uparrow_{\vec{r}_{2}}\right| +
\left|\downarrow_{\vec{r}_{1}}\downarrow_{\vec{r}_{2}}\right>
\left<\downarrow_{\vec{r}_{1}}\downarrow_{\vec{r}_{2}}\right|\right)
\\ & & +
\frac{1+\vec{r}_{1}\cdot\vec{r}_{2}}{4}\left(\left|\uparrow_{\vec{r}_{1}}\downarrow_{\vec{r}_{2}}\right>
\left<\uparrow_{\vec{r}_{1}}\downarrow_{\vec{r}_{2}}\right| +
\left|\downarrow_{\vec{r}_{1}}\uparrow_{\vec{r}_{2}}\right>
\left<\downarrow_{\vec{r}_{1}}\uparrow_{\vec{r}_{2}}\right|\right)
\end{array}.
\end{equation}
The whole density matrix is
\begin{equation}
\rho^{hvt} =
\prod_{\vec{r}_{1}\vec{r}_{2}}\otimes\rho_{\vec{r}_{1}\vec{r}_{2}},
\label{3dentangle}
\end{equation}
The reduced density matrix satisfies equ(\ref{correlate}) and
equ(\ref{single}), which is the content of QS-I and QS-II. For
QS-III and QS-IV, although we will skip the details here, a state
after measurement can be constructed easily. Building a state on the
equivalent classes solves the problem of QS-V. We have successfully
constructed a classical theory for two-spin quantum system. It is a
classical theory but it violates Bell's inequality. As we argued
above, we already know that, due to inconvenience and inconsistency,
this theory should not be preferred. However, we can still ask how
can such a classical theory does succeed to give all expected
results from QM? The answer is it includes non-local information.

In \cite{proof}, Bell's inequality was proved more generally with
only the locality assumption, their equ($2^{\prime}$) uses
\begin{equation}
p_{1,2}\left(\lambda, a, b\right) = p_{1}\left(\lambda,
a\right)p_{2}\left(\lambda, b\right), \label{locality}
\end{equation}
where $\lambda$ is a hidden variable independent of $a,b$ to express
the idea of measurement-independent reality of a quantum system. Since
our IHVT violates Bell's inequality, we want to check if it respects
the above equation. Consider the situation where we measure direction $a$ and
$b$ on those two spins respectively.
\begin{equation}
\begin{array}{lll}
\left<\hat{S}^{1}\hat{S}^{2}\right>\left(a,b\right) & = &
tr\left(\hat{S}^{1}\left(a\right)\hat{S}^{2}
\left(b\right)\rho\left(\vec{\lambda}\right)\right)
\\
& = & \sum_{\lambda_{ab}}
\left<\lambda_{ab}\right|\hat{S}^{1}\left(a\right)\hat{S}^{2}
\left(b\right)\left|\lambda_{ab}\right>f\left(\lambda_{ab}\right)
\\
& = & \sum_{\lambda_{ab}} s^{1}\left(a,\lambda_{ab}\right)s^{2}
\left(b,\lambda_{ab}\right)f\left(\lambda_{ab}\right)
\end{array}
\end{equation}
The left hand side can be regarded as
\begin{equation}
\left<\hat{S}^{1}\hat{S}^{2}\right>\left(a,b\right) =
\sum_{\lambda_{ab}}s^{1}s^{2}\left(a,b,\lambda_{ab}\right)f\left(\lambda_{ab}\right).
\end{equation}
From the core of the integral, we see that
\begin{equation}
s^{1}s^{2}\left(a,b,\lambda_{ab}\right)=s^{1}\left(a,\lambda_{ab}\right)s^{2}
\left(b,\lambda_{ab}\right),
\end{equation}
or generally,
\begin{equation}
s^{1}s^{2}\left(a,b,\vec{\lambda}\right)=s^{1}\left(a,\vec{\lambda}\right)s^{2}
\left(b,\vec{\lambda}\right). \label{nonlocality}
\end{equation}
Compared with equ(\ref{locality}), equ(\ref{nonlocality}) does look
like an expression of locality, with the difference that a single
hidden variable is replaced by many hidden variables. However, it is
this replacement that introduces non-local information, because the
effective one out of $\vec{\lambda}$ is $\lambda_{ab}$, which does
depend on both $a$ and $b$, the measurements on both spins. During
the measurement process, a sample should be drawn from an effective
probability distribution. And the effective one has to be determined
through information with both directions $a$ and $b$ together. It is
as if the system has to know both directions to make its decision.
It is definitely contextual.

We have explicitly shown the place non-locality comes into QM.
When the classical theory is used to describe QM, we have to
require non-local information. If this non-locality is
unacceptable, then we should rule out the idea of HVT. However,
this never means QM in its own language requires non-local
information. This is a topic which has never been addressed in this paper.

\section{Conclusion and Discussion}
\label{conclusion}

In a summary, to find a classical theory respecting all five QS
facts, our conclusion is: first, single variable HVT is incompatible
with non-commutative relation between operators; second, even if all
operators are commutative, the inherent relation between them has to
be abandoned; third, the exclusive-event HVT requires contextuality
between object and observer; fourth, both EHVT and IHVT suffer from
the inconsistency problem: the expression used to denote the state
is different with the one used to recover probability; and at last,
IHVT is shown to imply non-locality. We find the price is
unreasonably high: even after we accept the non-locality, CPT-III
need to be twisted. And due to those twists, such a classical system
could no longer be broadcasted. Noticing CPT-III is essential to
make it possible to broadcast a classical system. The possibility of
being broadcasted is one key fact in understanding of classical
measurement. Even theoretically, not depending on the experimental
test of Bell's inequality, the idea of HVT should be discarded from
theory of quantum systems.

In another words, under reasonable consideration it is impossible to
map a full-structure density matrix to a diagonal density matrix.
With this conclusion in mind, we may say that
although the current language of QM may not be the ultimate one,
any equivalent language should include existence of off-diagonal
elements of the density matrix and allow vectors to be transformed from
one basis to another, which is only possible when operators do not
always commute with each other. We know quantum
measurement is not equal to classical measurement
of a TRCO. Classical measurement creates a broadcast, but quantum
measurement does not.

Finally, we are not saying those are all the possibilities of CPT
for QS. From the $C^{*}$-algebra point of view, what we have tried here are just
two examples of multiplications between operators,
$\sigma_{\vec{r}_{1}}\sigma_{\vec{r}_{2}}=0$ for the exclusive case and
$\sigma_{\vec{r}_{1}}\otimes\sigma_{\vec{r}_{2}}$ for the independent
case. There could be some other kinds of algebras among operators.
If we assume symmetry among $\sigma$ operator on all directions,
them those two are the only choices. Besides our own HVTs we have
also examined Bell's HVT and Bohm's HVT and ruled them out based on
repeat measurement and validity of CPT-III.

\section{Acknowledgment}
Thanks should first be given to Dr. Leslie Ballentine and Dr. Ian
Affleck for their inspiring courses on quantum mechanics,
giving a detailed and insightful check of the basis of
quantum mechanics, and an fascinating introduction into the
fundamental problems in quantum mechanics. Thanks should also be
given to Dr. Mona Berciu for her comments and discussions on this
work, and Dr. Robert Peter and Janelle von Dongen for reading through the manuscript. At last, we
acknowledge the anonymous referee gratefully for the comments and
suggestions during the revision.


\begin{thebibliography}{99}
\bibitem{footnote}If we allow contextual communication between
object and observer then $Q$ could be equal to $1$. Or $Q$ could also be
equal to $1$ if our device is totally unsharp, say it reads positive
when $\vec{r}\cdot\vec{r}_{0}>0$ where $\vec{r}$ is the vector's
direction and $\vec{r}_{0}$ is device's direction. But the later is
not the case we are discussing and theoretically it is always
possible to make the device sharp.
\end{thebibliography}
\end{document}